\documentclass[twocolumn,prl,showpacs,floatfix]{revtex4}
\usepackage{graphicx}
\usepackage{bm}
\usepackage{amsmath}
\begin{document}
\title{Coulomb Blockade and Kondo Effect in a Quantum Hall Antidot}
\author{H.-S. Sim,$^{1}$ M. Kataoka,$^{2}$ Hangmo Yi,$^1$ N. Y. Hwang,$^3$
M.-S. Choi,$^3$ and S.-R. Eric Yang$^3$}
\affiliation{
$^1$School of Physics, Korea Institute for Advanced Study, 207-43
Cheongryangri-dong, Dongdaemun-gu, Seoul 130-722, Korea 
}
\affiliation{
$^2$Cavendish Laboratory, Madingley Road, Cambridge CB3 0HE, United Kingdom 
}
\affiliation{
$^3$Department of Physics, Korea University, 1 5-ka Anam-dong, Seoul 136-701,
Korea 
}
\date{\today}
\begin{abstract}
We propose a general capacitive model for an antidot, which has two localized
edge states with different spins in the quantum Hall regime.
The capacitive coupling of localized excess charges, 
which are generated around the antidot due to magnetic flux quantization,
and their effective spin fluctuation
can result in Coulomb blockade, $h/(2e)$ Aharonov-Bohm oscillations, and 
the Kondo effect. 
The resultant conductance is
in qualitative agreement with recent experimental data.
\end{abstract}

\pacs{73.23.Hk, 72.15.Qm, 73.43.-f}

\maketitle

The Kondo effect arises due to many-body interactions between a 
localized spin and free electrons \cite{Kondo,Hewson}.
Recently, there has been renewed interest in the effect 
as it was predicted \cite{Glazman,Ng} 
and observed \cite{Goldhaber-Gordon,Cronenwett,van_der_Wiel}
in quantum dots.
In a quantum dot, the localized spin is naturally provided
when the dot has an odd number of electrons.

\begin{figure}
\includegraphics[width=0.45\textwidth]{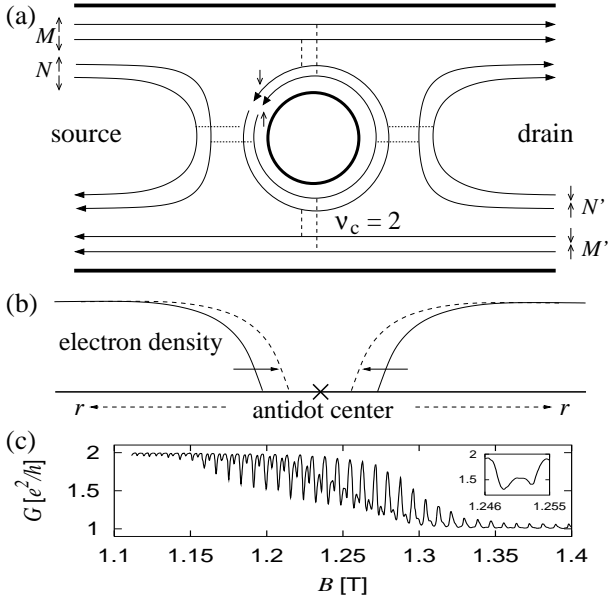}
\caption{
(a) A quantum Hall antidot
with antidot states and extended edge channels ($M,M',N,N'$) 
with their spins (solid arrows),
which come from the 1st (the antidot states, $M,M'$)
and the 2nd Landau levels ($N,N'$).
The intra and the inter-Landau level couplings are drawn
as dashed and dotted lines, respectively.
$B$ is applied perpendicularly to the plane.
(b) Schematic diagram of electron density shift (solid arrows) around
an antidot as $B$ increases; this generates local charge imbalance 
with respect to the background charge.
(c) An experimental result under 25 mK and zero bias
(see Ref. \cite{Masaya_Kondo}).
Inset: $G(B)$ for one AB period around $B = 1.25$ T.
}
\label{geometry}
\end{figure}

Quantum antidots in the integer quantum Hall regime have
attracted recent interest in connection with experimental
observations of the charging effect \cite{antidot_charging},
$h/(2e)$ Aharonov-Bohm (AB) oscillations \cite{antidot_charging,Masaya_double},
and Kondo-like signatures \cite{Masaya_Kondo}.
In these systems, localized quantum
Hall edge states are formed along an equipotential line of a
``potential hill'' which defines the antidot.
As in quantum dots, electrostatic interaction of the 
localized antidot states may give rise to the charging effect.
However, the magnetic flux quantization makes the antidots
rather intriguing.
When magnetic field $B$ changes adiabatically,
each single-particle state encircling the antidot moves
with respect to the antidot potential, 
adjusting the enclosed antidot area $S$
in order to keep the flux $BS$ constant [Fig. \ref{geometry}(b)]. 
This electron displacement results in charge imbalance around the antidot,
i.e., local accumulation of excess charge $\delta q(B)$
\cite{Masaya_double},
which is the source of the 
charging effect \cite{antidot_charging}. 
The accumulated $\delta q(B)$ is relaxed via 
single electron resonant tunnelings \cite{FQHE}.
These tunneling events occur
$\nu_c$ times within one AB period $\Delta B (=h/eS)$
when the antidot has $\nu_c$ localized edge states
\cite{Kim}.
Also $\delta q(B)$ is periodic with the period $\Delta B$.


The origin of the Kondo-like signature in the antidots
is not understood yet.
One may naively consider that
the spin-split single-particle antidot states support
a localized spin.
However, their SU(2) spin symmetry may be broken by the Zeeman energy
and thus they can not cause the signature.
Rather, many-body antidot states may play an important role.

In this Letter, we provide a theoretical model for
the Kondo effect in a quantum Hall antidot system.
As a natural way to incorporate the charging effect,
capacitive interactions of excess charges with different spins
are adopted, and the source-drain conductance $G(B)$ is computed within 
the model.
We find that {\em the effective spin flips of the excess charges} can
cause the Kondo effect.
Within one AB period $\Delta B$, $G(B)$ can show approximately
two normal resonances and one Kondo resonance,
consistent with experimental data \cite{Masaya_Kondo}.
The two normal resonant tunneling events, involving spin-down electrons,
are evenly spaced with varying $B$,
constituting $h/(2e)$ AB oscillations, 
while the normal tunneling of spin-up electrons is Coulomb blockaded.
These characteristic features of the resonances result from
{\em the interaction between the excess charges with different spins},
and they
can be tested by measuring conductance and higher cumulants of counting 
statistics such as shot noise.

Figure \ref{geometry} shows the 
experimental setup and data in Ref. \cite{Masaya_Kondo}.  
The data show a $2e^2/h$ plateau ($\nu_c \sim 2$) of $G(B)$ with
a series of dips,
which consist, within one AB period, of
two normal AB resonance dips and
one intermediate Kondo-like dip
[see, eg., the data around $B = 1.25$ T 
in the inset of Fig. \ref{geometry}(c)].
These observations are consistent with the assumption that 
the antidot states come only from the
two spin-split branches of the lowest Landau-level (LL)
and that $\Delta B \ll B$.
The system has both 
the localized antidot states and
the extended edge states.
The antidot hamiltonian
can be written as
$H_\mathrm{AD} = \sum_{m\sigma} (\hbar \omega_c/2 + E^Z_\sigma + V_m)
c^\dagger_{m\sigma} c_{m\sigma}
+ \sum_{m m' n n' \sigma \sigma'}
W_{m m' n n'} c_{n' \sigma'}^\dagger c_{n \sigma}^\dagger
c_{m \sigma} c_{m' \sigma'}$,
where
$\omega_c$ is the cyclotron frequency, $E^Z_\sigma$ is the Zeeman energy,
$V$ is the antidot potential energy
(including a positive background term preserving  
total charge neutrality),
$W$ is the Coulomb interaction,
and
$c_{m \sigma}^\dagger$ 
creates an
electron with spin $\sigma$
in a localized single-particle state 
(enclosing $m$ magnetic flux quanta).
The hamiltonian of the extended edge states is
$H_{\rm edge} = \sum_{ik\sigma} \epsilon_{ik\sigma}
c^\dagger_{ik\sigma} c_{ik\sigma}$, where
$c^\dagger_{ik\sigma}$ create
an electron 
in an extended state
(with momentum $k$, energy $\epsilon_{ik\sigma}$, 
spin $\sigma$, 
and index $i$).
We only consider
extended states coming from the first and the second LLs
($i \in \{M, M', N, N' \}$)
[Fig. \ref{geometry}(a)];
the extended states of higher LLs have negligibly small
tunneling amplitudes $V^i_{km\sigma}$ to antidot.
The total hamiltonian of the system is
$H_{\rm tot} = H_\mathrm{AD} + H_{\rm edge} + H_{\rm tun}$
with $H_{\rm tun} = \sum_{ikm\sigma} V^i_{km\sigma}
c^\dagger_{ik\sigma} c_{m\sigma} + \text{H.c.}$

The energy $E_\mathrm{AD}$ of the antidot Hamiltonian $H_\mathrm{AD}$
varies
with $B$ due to the formation [Fig. \ref{geometry}(b)]
of localized excess charges $\delta q_\sigma(B)$.
Since 
$\delta q_\uparrow$ and $\delta q_\downarrow$ 
are spatially separated from each other and
from extended edge states by incompressible regions,
one can approximate $E_\mathrm{AD}$ using an
{\em effective} capacitance matrix $C$ as:
\begin{eqnarray}
E_\mathrm{AD}(\delta q_\uparrow(B), \delta q_\downarrow(B)) 
= \frac{1}{2} \sum_{\sigma \sigma'}
\delta q_\sigma \left(C^{-1}\right)_{\sigma \sigma'} \delta q_{\sigma'}.
\label{effH}
\end{eqnarray}
$\delta q_\sigma$ can have a form \cite{deltaq}
of
$\delta q_\sigma = -eN_\sigma^{\rm AD} - Q_{G\sigma} - Q_{B\sigma}(B)$,
where $N_\sigma^{\rm AD}$ is the total number of electrons with spin
$\sigma$ occupying the antidot state, $Q_{G}$ is the ``antidot-gate''
charge (independent of $B$), and $Q_B(B)$ gives the dependence on $B$.
$C_{\sigma \sigma'}$ 
is a classical electrostatic
quantity if alternating compressible and incompressible regions 
\cite{Chklovskii}
are formed around the antidot.
In other cases, it is a phenomenological parameter.
We have 
$|C_{\uparrow \downarrow}| < C_{\uparrow \uparrow} < 
C_{\downarrow \downarrow}$,
since $\delta q_\downarrow$ 
is located outer from antidot center than
$\delta q_\uparrow$
due to its higher $E^Z_\sigma$;
$|C_{\uparrow \downarrow}|$ is the smallest as it is 
a mutual capacitance.
Because $C_{\sigma \sigma'}$ varies very slowly 
within one AB period,
we will take it as constant.
Note that $E^Z_\sigma$
can be counted in the definitions of $\delta q_\sigma$ and $C$,
as it also causes the displacement and
the separation of $\delta q_\sigma$'s.
Equation (\ref{effH}) is a good approximation for large-size antidots
with $B \gg \Delta B$,
and it is analogous to 
the constant interaction (CI) model of quantum dots \cite{Glazman_CI}.
As in the CI model,
the validity of Eq. (\ref{effH}) is tested within a Hartree-Fock
approximation.
Some terms independent of $N^{\rm AD}_\sigma$ have been ignored in
Eq. (\ref{effH}) just like in the CI model.
The control of $\delta q_\sigma$ by $B$
is reminiscent of
the charge control by gate voltage in quantum dots.

Coulomb blockade prohibits tunneling unless one of the
following conditions is satisfied.  First, a
normal resonance occurs whenever the antidot Hamiltonian is
invariant under a single electron tunneling, i.e.,
\begin{align}
E_\mathrm{AD}(\delta q_\uparrow \pm e, \delta q_\downarrow) = 
E_\mathrm{AD}(\delta q_\uparrow, \delta q_\downarrow),
\label{resonance_condition_up} \\
E_\mathrm{AD}(\delta q_\uparrow, \delta q_\downarrow \pm e) = 
E_\mathrm{AD}(\delta q_\uparrow, \delta q_\downarrow).
\label{resonance_condition_down}
\end{align}
Another allowed tunneling process is through
a Kondo resonance, in which an electron tunnels
into the antidot states and another electron
with the opposite spin tunnels out via a virtual state.
The condition for this is
\begin{equation}
E_\mathrm{AD}(\delta q_\uparrow \pm e, \delta q_\downarrow \mp e) = 
E_\mathrm{AD}(\delta q_\uparrow, \delta q_\downarrow).
\label{resonance_condition_Kondo} 
\end{equation}
Among many available virtual states, the one with
the lowest energy (thus with the biggest contribution) is
either $(\delta q_\uparrow \pm e, \delta q_\downarrow)$ or
$(\delta q_\uparrow, \delta q_\downarrow \mp e)$.

\begin{figure}
\includegraphics[width=0.5\textwidth]{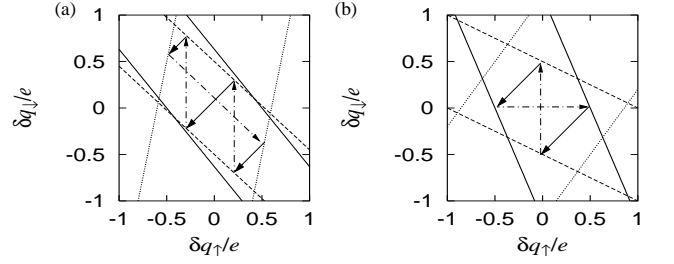}
\caption{
($\delta q_\uparrow$, $\delta q_\downarrow$) evolves along
a solid-arrow trajectory as $B$ increases. 
When one of resonance conditions is met,
($\delta q_\uparrow$, $\delta q_\downarrow$) jumps following
horizontal (normal spin-up electron tunneling), vertical (spin down), or 
diagonal dashed-dot arrows (Kondo resonance).
Solid, dashed, and dotted lines indicate resonance conditions 
of
Eqs.\ (\ref{resonance_condition_up}),
(\ref{resonance_condition_down}), and
(\ref{resonance_condition_Kondo}), respectively.
Two different parameters are used: 
(a) $\alpha \equiv |C_{\uparrow \downarrow}|/C_{\uparrow \uparrow} 
= 0.95$ and (b) $0.5$.
For both the cases, we choose 
$C_{\downarrow\downarrow}/C_{\uparrow\uparrow}=1.2$.
}
\label{trajectory}
\end{figure}

Assuming the antidot potential varies linearly
within the scale of the magnetic length $l_B(\equiv\sqrt{h/eB})$,
$\delta q_\sigma$ may be approximated 
as a linear function of $B$
between two adjacent tunneling (i.e., relaxation) events.
Since
the spatial separation of $\delta q_\uparrow$ and $\delta q_\downarrow$
is small compared with their average radius, 
$d\delta q_\uparrow/dB \simeq d\delta q_\downarrow/dB$.
Therefore, as $B$ increases, the ground state value of
($\delta q_\uparrow$,$\delta q_\downarrow$) evolves
parallel to the line of $\delta q_\uparrow = \delta q_\downarrow$
until it jumps at one of the resonance conditions of Eqs. 
(\ref{resonance_condition_up},\ref{resonance_condition_down},\ref{resonance_condition_Kondo}) [see Fig.\ \ref{trajectory}].
The evolution trajectory of $(\delta q_\uparrow, \delta q_\downarrow)$ is 
periodic with period $\Delta B$.
Depending on a
parameter $\alpha \equiv |C_{\uparrow \downarrow}|/C_{\uparrow \uparrow}$
and initial values of $(\delta q_\uparrow, \delta q_\downarrow)$,
two different types of 
trajectories are possible when
$C_{\uparrow \uparrow} < C_{\downarrow \downarrow}$:
(i) two evenly spaced consecutive 
tunnelings of spin-down electrons as well as
one intermediate Kondo resonance
[Fig.\ \ref{trajectory}(a)] or (ii) alternating tunnelings of
spin-up and down electrons with an arbitrary phase difference
[Fig.\ \ref{trajectory}(b)].  If $\alpha=1$
(maximum), all trajectories are of type (i)
regardless of the initial value of $\delta q_\sigma$.  In the other
extreme case $\alpha=0$ (minimum), i.e., $C_{\uparrow \downarrow}=0$, 
the two spin states are completely decoupled and all trajectories are
of type (ii).  For $0<\alpha<1$, both types are allowed 
depending on initial $\delta q_\sigma$.
For larger $\alpha$, more trajectories follow the type (i).
We remark that the experimental data \cite{Masaya_Kondo}
are consistent with the type (i):  At high temperature,
they show $h/(2e)$ AB oscillations, i.e.,
evenly spaced dips of $G(B)$ which appear two times per $\Delta B$.
As the temperature
is lowered, another dip appears approximately
once every $\Delta B$.  We argue that the experiment
is in the parameter regime where type (i) trajectories are
prevalent, 
i.e., $\alpha\sim1$.  
The scale that separates the high and low temperature regimes
is the Kondo temperature $T_K$. 

We now describe in detail
how the Kondo effect arises in our model.
In the vicinity of a Kondo resonance condition,
only the two lowest
[$(\delta q_\uparrow,\delta q_\downarrow)$ and
$(\delta q_\uparrow + e,\delta q_\downarrow -e)$]
and the next two excited states
[$(\delta q_\uparrow+e,\delta q_\downarrow)$ 
and $(\delta q_\uparrow,\delta q_\downarrow-e)$]
of $H_\textrm{AD}$ are important;
we ignore all the other excited states,
which may affect the Kondo effect only
slightly as in multilevel quantum dots \cite{multilevel}.
Using these four states one can map $H_\textrm{AD}$ into
the Anderson impurity model, given by a truncated hamiltonian
$H_{\rm imp}
= \sum_\sigma \epsilon_\sigma d_\sigma^\dagger d_\sigma + 
U d_\uparrow^\dagger d_\downarrow^\dagger d_\downarrow d_\uparrow$,
where $d_\sigma^\dagger$ creates an electron 
in the impurity site.
The two lowest states constitute the two
singly occupied impurity states,
while the next two excited states 
do the empty and doubly occupied states.
Defining 
$E_\mathrm{empty}
\equiv E_\mathrm{AD}(\delta q_\uparrow+e,\delta q_\downarrow)$, 
we get $\epsilon_\uparrow
= E_\mathrm{AD}(\delta q_\uparrow,\delta q_\downarrow)-E_\mathrm{empty}$, 
$\epsilon_\downarrow =
E_\mathrm{AD}(\delta q_\uparrow+e,\delta q_\downarrow-e)-E_\mathrm{empty}$, 
and $U+\epsilon_\uparrow+\epsilon_\downarrow
= E_\mathrm{AD}(\delta q_\uparrow,\delta q_\downarrow-e)-E_\mathrm{empty}$.  
Here,
$\Delta \epsilon \equiv \epsilon_\uparrow - \epsilon_\downarrow$ behaves
as an {\em effective} Zeeman energy of the impurity site;
$\Delta \epsilon = 0$ at the Kondo resonance condition.
$H_{\rm tot}$ can now be mapped into the impurity coupled to
the extended edge states by amplitude $V^i_{k\sigma}$:
\begin{eqnarray}
H_{\rm K} = 
H_{\rm imp} 
+ H_{\rm edge} + 
\sum_{ik\sigma} V^i_{k\sigma}
c^\dagger_{ik\sigma} d_\sigma + \text{H.c.}
\label{Anderson}
\end{eqnarray}

One can estimate energy scales from experimental data.
Since only down-spin electrons cause the normal resonances
in the type (i) trajectories, 
we rewrite Eq. (\ref{effH}) as
$E_\textrm{AD}(\delta q_\uparrow, \delta q_\downarrow)
= (\delta q_\downarrow + \alpha \delta q_\uparrow)^2/(2C_{\rm out}) +
\delta q_\uparrow^2/(2C_{\rm in})$, where
$C_{\rm out} = C_{\downarrow \downarrow} - \alpha |C_{\uparrow \downarrow}|$
and $C_{\rm in} = C_{\uparrow \uparrow}$
\cite{IRCR}.
One can then easily see that
$e^2/C_{\rm out}$ corresponds to the charging energy ($\sim 60$ $\mu eV$) 
measured in Ref. \cite{Masaya_Kondo}.
Note that 
$- \epsilon_\sigma$, $U + \epsilon_\sigma \sim e^2/(2C_{\rm out})$.
When the effective Zeeman energy $\Delta \epsilon$ has a finite value,
the Kondo resonance is split and suppressed \cite{Meir2,resolution},
and thus the Kondo signature of $G(B)$ can only appear within
a certain range $\delta B$ in one AB period $\Delta B$.
For $\alpha = 1$, 
the Kondo resonance occurs at
$\delta q_\uparrow = \pm e/2$,
and in its vicinity where
$\delta q_\uparrow = \pm (1/2 - p)e$,
we get
$\Delta \epsilon = p e^2/C_{\rm in}$.
From this, one can roughly estimate that
$e^2/C_{\rm in}$ is of the order of
10 $\mu eV$ in the experimental situation \cite{Masaya_Kondo},
where
$\Delta \epsilon \lesssim$ 10 $\mu eV$
(the energy scale of zero bias anomaly), 
the width $\delta B$ of the Kondo dip $\simeq \Delta B / 4$,
and $p \simeq \delta B / \Delta B$.

The resonance width
$\Gamma_\sigma(E) \equiv \sum_i \Gamma^i_\sigma(E)$
has a spin dependence
since $V^{i=M(M')}_{k\uparrow} \neq V^{i=M(M')}_{k\downarrow}$
(Fig. \ref{geometry}),
where $\Gamma^i_\sigma(E) = 2 \pi \sum_k
|V^i_{k\sigma}|^2 \delta (E - \epsilon_{ik\sigma})$.
We will ignore $k$ dependence of $V^i_{k\sigma}$ 
for simplicity.
The spin-dependent $\Gamma_\sigma$'s renormalize
the effective Zeeman energy as
$\Delta \tilde{\epsilon} (\delta q_\uparrow, \delta q_\downarrow)$,
as in the 
quantum dots coupled to
ferromagnetic leads \cite{Zhang}.
This changes the Kondo resonance
condition to 
\begin{eqnarray}
E_\mathrm{AD}(\delta q_\uparrow \pm e, \delta q_\downarrow \mp e) = 
E_\mathrm{AD}(\delta q_\uparrow, \delta q_\downarrow) 
+ \Delta \tilde{\epsilon},
\label{redefined}
\end{eqnarray}
instead of Eq. (\ref{resonance_condition_Kondo}).
When Eq. (\ref{redefined}) is satisfied, 
there is no renormalized effective Zeeman splitting so that 
one can estimate \cite{Haldane}
$T_K \sim (\sqrt{\Gamma U}/2) \exp(\pi \epsilon (U+\epsilon)/\Gamma U)$,
where $\Gamma = (\Gamma_\uparrow + \Gamma_\downarrow)/2$.
By using $\alpha=1$ and
estimating energy scales \cite{estimation}, 
we perform the numerical renormalization group (NRG) calculation 
\cite{NRG,Hewson}
and find $T_K \sim 1$ $\mu eV$, 
which has 
the same order of magnitude
with the zero-bias anomaly of the Kondo-like data
\cite{Masaya_Kondo}.

\begin{figure}
\includegraphics[width=0.4\textwidth]{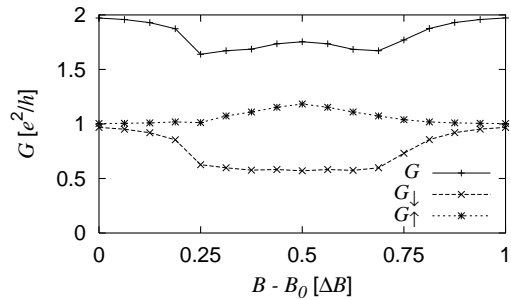}
\caption{
Calculated $G(B)$ and $G_\sigma(B)$ 
around a magnetic field $B_0$ ($ \sim 1$ T). 
$G(B)$ matches well the experimental data, eg.,
the inset in Fig.\ \ref{geometry}(c).
}
\label{NRGconductance}
\end{figure}

Now we derive the conductance 
at zero temperature and zero bias
\cite{Ng,Kirczenow}
as $G(B) = \sum_\sigma G_\sigma(B)$ where
$G_\sigma = (e^2/h) (1 + T_\sigma \sin^2 \theta_\sigma)$.
Here, $T_\sigma \equiv 4(\Gamma^{i=N}_\sigma \Gamma^{i=N'}_\sigma
- \Gamma^{i=M}_\sigma \Gamma^{i=M'}_\sigma)/\Gamma_\sigma^2$ and
$\theta_\sigma \equiv \textrm{Im} \ln {\cal G}_\sigma(E + i\delta)$,
where ${\cal G}_\sigma(E)$ is the Green's function of the impurity site.
$\Gamma^i_\sigma$'s can be estimated
\cite{estimation}
from the experiments in the following way:
Since $G(B)$ shows dips below $2e^2/h$ [Fig.\ \ref{geometry}(c)]
and the excess charges follow a type (i) trajectory, we expect
$\Gamma^{i=M,M'}_\downarrow > \Gamma^{i=N,N'}_\downarrow$ and
$T_\downarrow < 0$, i.e.,
the down spin electrons favor the backward scattering.
On the other hand, 
the $\Delta B$ periodic peaks at stronger magnetic fields near 
the plateau of $G \simeq e^2/h$ 
may be interpreted 
as normal resonances of spin-up 
electrons, 
and $\Gamma^{i=M,M'}_\sigma$ becomes smaller as $B$ decreases.
One thus obtains that
$\Gamma^{i=M,M'}_\uparrow < \Gamma^{i=N,N'}_\uparrow$ and
$T_\uparrow > 0$ for weaker $B$ where $G > e^2/h$.
Then we can calculate $G(B)$ for an AB period
by using the NRG results of $\sin^2 \theta_\sigma$
obtained for $\alpha=1$.
The result (see Fig.\ \ref{NRGconductance})
is in qualitatively good agreement with the experimental data,
eg., the inset of Fig.\ \ref{geometry}(c).

Figure \ref{NRGconductance} shows a spin-dependent behavior of $G_\sigma(B)$.
$G_\uparrow$ has a Kondo peak,
while $G_\downarrow$ shows a mixture of two normal
and one intermediate Kondo resonance dips.
As a result,
at the center of the Kondo resonance,
even if the unitary Kondo limit is reached,
$G$ can have a larger value than $e^2/h$.
The center and the width of the Kondo resonance in $G_\sigma(B)$ is
governed by Eq. (\ref{redefined}) and $e^2/C_{\rm in}$, respectively.
This spin-dependent behavior 
results from 
the type (i) trajectory and 
multiple extended edge states (i.e., the sign of $T_\sigma$),
and is an interesting feature of the antidot Kondo effect.
Our predictions on $G_\sigma(B)$
may be tested by measuring conductance and 
higher cumulants of counting statistics such as shot noise \cite{Blanter},
which can resolve $G$ into $G_\uparrow$ and $G_\downarrow$.

In the broad range of many AB periods,
as $B$ increases,
the spin-down antidot state becomes more strongly coupled 
to the extended 
states with $i=M,M'$ and then eventually disappears.
Then, $G(B)$ decreases from $2e^2/h$ to $e^2/h$,
and the Kondo effect may become
enhanced at first and then suppressed
around $B$ where $G \simeq e^2/h$.
Note that our model can also naturally explain $h/(2e)$ AB oscillations 
without any Kondo signature at higher $B \sim$ 3T \cite{Masaya_double} 
where the spin-up antidot state is decoupled with 
all extended edge states.

Finally, we note that 
as a Kondo resonance is crossed by increasing $B$,
the total electron spin of the ground state {\it decreases}
[see Fig.\ \ref{trajectory}(a)].
To understand the microscopic origin of this behavior, 
we perform a Hartree-Fock (HF) calculation \cite{Yang}, in which
the electron antidot system is transformed
into a system confining holes
by the mapping  
$c_{m\sigma} \rightarrow h_{m\sigma}^\dagger$ and
$c_{m\sigma}^\dagger \rightarrow h_{m\sigma}$. 
Here, $h_{m\sigma}^\dagger$ creates a hole.
We use an inverse bell-shape confinement potential such as
$a r^2$ for $r < r_c$ and $b + c/r^2$ for $r > r_c$,
so that hole states near the edge are affected by the $1/r^2$
potential.
Our HF calculation with about 50 confined holes shows that 
there appear the ground state transitions from
$| N_{\downarrow},N_{\uparrow} \rangle$ to 
$ | N_{\downarrow}-1,N_{\uparrow}+1 \rangle$
with increasing $B$
when the curvature of the 
confinement potential near the edge of the system is negative.
$| N_{\downarrow}, N_{\uparrow} \rangle$ 
is a maximum density droplet state \cite{Yang} with 
$N_{\uparrow}$ spin-up and $N_{\downarrow}$ spin-down holes
($N_{\downarrow}>N_{\uparrow}$ due to the Zeeman energy)
and the droplet size
is $\sim \sqrt{2N_\downarrow} l_B$.
The transitions, 
where the total electron spin decreases, 
occur because
for the negative-curvature potential, 
the total confinement energy of holes
increases faster than the total Coulomb energy as $B$ increases.
This causes the hole droplet to minimize its size.
Although the HF calculation with a relatively small
number of holes can not be directly compared with large-size antidots,
it demonstrates 
the existence of the spin-flip ground state transition.


To conclude, we have constructed a capacitive model which can describe 
many-body antidot states in integer quantum Hall regimes, 
and predicted Coulomb blockade and
characteristic AB oscillations with the Kondo effect.
Our model may be applicable to 
large-size quantum dots or rings 
in $\nu = 2$ quantum Hall regimes \cite{Staring} as well. 


HSS and MK thank C. J. B. Ford and H. Schomerus 
for helpful discussions and their hospitality
in Cambridge and Dresden.
The discussions with K. Kikoin, K. Kang, and T. S. Kim are 
gratefully acknowledged.
We were supported by SKORE-A (HY and MSC), eSSC (MSC),
and QSRC at Dongguk University (SREY).

\bibliographystyle{apsrev}

\begin{thebibliography}{99}
\bibitem{deltaq} 
If only incompressible regions (or maximum-density droplets of holes)
are formed around the antidot,
one estimates
$\delta q_\sigma \simeq (e n_e) \delta S$
in terms of electron density $n_e = eB/h$ per spin
and area shift $\delta S$ caused by 
field change $\delta B \equiv B-B_0$ ($< \Delta B$) from a given $B_0$.
Using $\delta (B S) = 0$, one finds
$Q_{B\sigma}(B) = Q_{B\sigma}(B_0) - e(B-B_0) / \Delta B$.
This estimation can be modified due to
charge redistribution,
if compressible regions are also formed.
\bibitem{IRCR}
When the antidot states are
the alternating compressible and incompressible regions,
$\delta q_\uparrow$ in the inner compressible region (CR)
can be screened by the outer CR
with a screening factor $\eta$
(for perfect screening, $\eta = 1$).
Then charges of $\delta q_\uparrow$ and
$\delta q_\downarrow + \eta \delta q_\uparrow$
are accumulated at the inner and outer CRs, 
respectively \cite{Masaya_double}.
They can be described by the rewritten $E_\textrm{AD}$
with $C_{\rm out}$, $C_{\rm in}$, and $\alpha \equiv \eta$.
\bibitem{resolution} 
In Ref. \cite{Masaya_Kondo},
the resolution is not good enough to see the splitting of Kondo signatures
at a finite bias.
\bibitem{estimation} 
We estimate $e^2/C_{\rm out} = 60$ $\mu eV$,
$e^2/C_{\rm in} = 12$ $\mu eV$,
and $\Gamma_\downarrow = 22$ $\mu eV$ (full width of half maximum) 
from the experimental data \cite{Masaya_Kondo}.
To obtain the ratios of $\Gamma^i_\sigma$'s
[$\propto \exp (-2 l_{i\sigma}/l_B$)],
we also crudely estimate hopping distances $l_{i\sigma}$ of the antidot
state with spin $\sigma$ to extended edge channels with index $i$ as
$l_{M\downarrow}= l_B$,
$l_{M\uparrow}= 1.7l_B$,
$l_{N\uparrow}= l_{N\downarrow}= 1.5l_B$, and
$l_{M'(N')\sigma}= l_{M(N)\sigma}$.
Note that $l_B \sim 25$ $nm$ when $B \sim 1$ T.
\end{thebibliography}


\end{document}